%% file: main.tex
\documentclass[sigconf]{acmart} % V1.2
\usepackage{verbatimbox}

\acmVolume{}
\acmNumber{}
\acmYear{}
\acmMonth{}
\acmDOI{}

\acmISBN{}

\acmConference[FDG2017]{Foundations of Digital Games Conference}{August 2017}{Cape Cod, MA} 
\begin{document}

% \markboth{Author Name}{Author Title}

\title{Deriving Quests from Open World Mechanics} 

\author{Ryan Alexander} % (rjalexan@ncsu.edu)
\affiliation{North Carolina State University}
\email{rjalexan@ncsu.edu}

\author{Chris Martens}
\affiliation{North Carolina State University}
\email{martens@csc.ncsu.edu}

\begin{abstract}

\input{abstract}

\end{abstract}


%\acmformat{}

\maketitle

%\begin{bottomstuff}

%\end{bottomstuff}

\section{Introduction}
\label{sec:intro}

\input{intro}

\section{Related Work}
\label{sec:relatedwork}
\input{related}

\section{Background}
\label{sec:background}
\input{background}

\section{A Formal Model of Minecraft}
\label{sec:model}
\input{model}

\section{Trace Analysis}
\label{sec:traceanalysis}
\input{tanalysis}

\section{Static Analysis}
\label{sec:staticanalysis}
\input{sanalysis}

\section{Conclusion}
\label{sec:conclusion}
\input{conclusion}

\appendix

% \begin{acks}
% \end{acks}

% Bibliography
\bibliographystyle{ACM-Reference-Format}
\bibliography{main}

\end{document}

%% file: abstract.tex
Open world games present players with more freedom than games with linear progression structures. However, without clearly-defined objectives, they often leave players without a sense of purpose. Most of the time, quests and objectives are hand-authored and overlaid atop an open world's mechanics.
But what if they could be generated organically from the gameplay itself? The goal of our project was to develop a model of the mechanics in Minecraft that could be used to determine the ideal placement of objectives in an open world setting. We formalized the game logic of Minecraft in terms of logical rules that can be manipulated in two ways: they may be executed to generate graphs representative of the player experience when playing an open world game with little developer direction; and they may be statically analyzed to determine dependency orderings, feedback loops, and bottlenecks. These analyses may then be used to place achievements on gameplay actions algorithmically.

%% file: intro.tex
In an {\em open world game}, players enjoy a great deal of autonomy in selecting from a complex system of mechanics, explorable spaces, and goals. Rather than experiencing a prescribed linear (or partially-ordered) progression of challenges and plot points orchestrated by a game designer, players are free to experiment with the consequences of the game's mechanics and exhibit more creativity in deciding what to do, frequently devising their own goals. In the framework of Jesper Juul \cite{juul2005hr}, open world games emphasize {\em emergence} over {\em progression}: they are more interested in opening a wide space of explorable consequences of the game's mechanics than in delivering a specific sequential experience.

Minecraft is a popular example of such a game, whose open world mechanics are supported by procedurally generated terrain: players can {\em mine} the terrain for natural resources, which they then {\em craft} into new tools necessary for survival, exploration, and additional mining. Minecraft also comes with a hand-authored {\em achievement tree} (see Figure 1) that serves to overlay some progression structure on top of the open world mechanics: the player is notified {\em Achievement Unlocked} when they reach certain points in their exploration of the game's mechanics, and eventually they may be led down any of several branches to find special items and characters.

Clearly, although open worlds without prescriptive goals have wide appeal, progression structures such as quests provide something that players are missing: a sense of purpose or progress as they navigate the game's mechanics. However, once players complete Minecraft's hand-authored achievement tree and special quests, they are back to where they started in a directionless open world. This quandary leads us to our problem statement: could we procedurally generate quests based purely on an open world game's mechanics?

Examination of Minecraft's hand-authored achievement tree suggests that much of it could be derived algorithmically. The tool crafting portion of the tree is a prime example of this, due to the cyclical nature of the crafting system. Making new tools and using new materials for tools are rewarded with achievements. By examining these gameplay elements we can determine why they were chosen to have achievements and how to apply that decision making to other elements.

We present a formal computational model of Minecraft's mining and crafting rules that is amenable to algorithmic treatment analogous to a {\em topological sort} of these mechanics. The results align naturally to progression structures in the game and suggest locations in game traces where achievements might be placed. Furthermore, we show how two distinct analyses of the rules provide distinct progression structures.

Our results constitute a novel approach to quest generation, based not on a model of story distinct from a game's mechanics, but in fact derived from them and directly leveraging their emergent behavior. As a secondary contribution, we illustrate the methodological value in using a formal modeling tool based on logic to gain insight into a game's mechanics.

%% file: related.tex
On a broad level, the context for this work is the field of procedural content generation (PCG). Within this field, we observe a shift in interest from algorithms that can create artbitrary, varied content to those that can generate {\em to fit a specfication}, such as the argument for generating with ASP~\cite{smith2012case}, a framework whose primary edge is its rich constraint specification capabilities.

Others have also observed that, while PCG traditionally has centered on crafting content to fit with the {\em emergent} dynamics of a game (such as level generation), generating progression structures such as narrative and quests poses novel and important challenges. For instance, the Grail framework~\cite{sullivan2012making} addresses the problem of constructing quests whose goals may have emergent solutions based on the game's mechanics, rather than scripted solutions based on the author's intended path for the player. 
Likewise, the Symon project~\cite{fernandez2012procedural} procedurally generates fetchquest-style narrative puzzles from a flexible definition of object interactions. Compared to our work, the goals of these projects were different: rather than starting with an open world and building quest content atop it, they aim to allow an author to specify only enough world rules for the generator to create a flexible range of playable story content.

The ``mission/space'' dichotomy~\cite{dormans2011generating} explicitly formulates the integration of generating spatial content (e.g. levels) with progression content (e.g. missions). First author Dormans also provides a more thorough theoretical account of integrating emergence and progression~\cite{dormans2011integrating}. Dormans et al.'s work closely relates to ours in that it explicitly addresses co-generation of an open, explorable world, and an ordered, finite progression, including consideration of how to generate a space to fit a provided mission, or an emergent structure from a progression. Our work explores the opposite direction, generating progression structures from emergent ones, and further considers richer world mechanics (crafting, mining) than the lock-and-key-based level exploration primarily explored by Dormans et al.

Recent interest in {\em general level generation}~\cite{khalifa2016general} poses a problem quite similar to the one we are addressing: how can a program, given a game's mechanics (specified in VGDL), generate a level (including a goal condition) for the game? In this framework, the level serves as the progression structure (an initial condition and a goal) whereas the games {\em rules}, specified in terms of which entities exist, their properties, and how they interact with other properties, create emergence. This problem aligns very well with the case we examine of taking Minecraft's open world rules and generating a beginning and end condition for a player to interact with them.

Finally, others have considered the problem of {\em the quest in a generated world}~\cite{ashmore2007quest}, which is similar to the idea of generating quests for an {\em open} world in that what a player can do within the world is not known at the outset. They formulate a quest in terms of spatial progression, and their system considers randomly generated levels and generates lock-and-key puzzles for them. In this work, the actions that make up the quest (collecting keys and unlocking doors) are layered atop a generated world without a pre-existing set of emergent rules. Our work seeks to construct quests not by adding affordances (keys, locks) to a generated world but by making use of existing game world rules, which permits us to directly integrate the actions a player takes to complete a quest with the mechanics that permit open play.

%% file: background.tex
\subsection{Emergence and Progression in Minecraft}
Emergence and progression are terms used by Jesper Juul~\cite{juul2005hr} to describe two different modes of gameplay. Progression structures engage the player through a structured set of objectives or quests. Emergent play revolves around the naturally occurring events that stem from a small number of simple rules. There are drawbacks to both methodologies used in isolation: games of pure progression may feel ``railroaded'' to the player, and furthermore demand more time from developers to hand craft the desired experience, while games of pure emergence risk overwhelming players with the vast space of possibilities available to them, or alternatively boring them with no clear direction for advancement. Thus, many games, such as Minecraft, utilize elements of both: they provide a simple set of open-ended mechanics for exploratory play while also adding quests and achievements to give the player a sense of progression.

Minecraft is an open world survival game, where players can traverse an infinite terrain of blocks. These blocks come in a variety of materials and can be collected by the player then used to build structures or make items. Due to the open world setting, there are no restrictions on where the player can go, which lends more freedom to the playing experience. Alongside this lack of boundaries, also comes a lack of explanation. Upon spawning into the game, players have ``no clear idea of what he or she can do within the world'' \cite{duncan2011}. The only hint is a message informing players that they have unlocked an achievement upon opening their inventory. These achievements are the main source of progression in this mostly emergent game.

\begin{figure}
	\includegraphics[width =\linewidth]{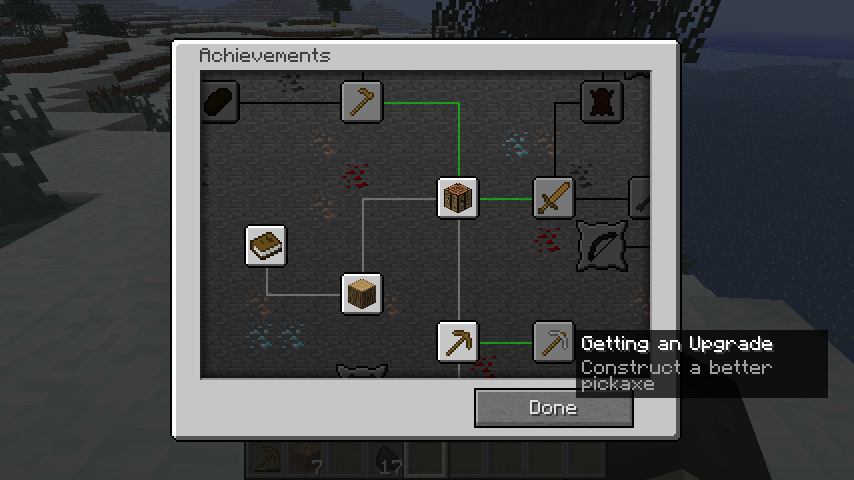}
	\caption{The achievement menu. Hovering the cursor over the icon for a visible achievement shows a brief description.}
	\label{fig:achieve}
\end{figure}

In Minecraft, achievements are optional objectives that the player can choose to accomplish. The achievement menu displays them as a tree where one achievement will serve as a requirement for completing another. There is no in-game reward for completing an achievement, other than unlocking the ability to see the next achievements in the tree. As shown in Figure \ref{fig:achieve}, each achievement has a square icon linked to other achievements by a line. The colors of the icons and lines represent the status of the achievement. Bright icons indicate that the achievement has been unlocked, darker icons mean that the achievement is not unlocked but can be unlocked and its details are visible to the player. The darkest icons signal that an achievement cannot be unlocked yet and the player does not have access to information about the achievement. Similarly, gray lines link unlocked achievements, green join unlocked achievements to those that can be unlocked, while black lines connect locked objectives.

\subsection{Quests, Achievements, and Objectives}
So far, we have been using the terms {\em quest}, {\em achievement system}, and {\em objective placement} relatively interchangeably. In this work, we adopt a simplified notion of {\em quest} meaning, essentially, a subset of the nodes in a gameplay trace that are recognized by the game system as progress (e.g. with ``achievement'' messages). An {\em objective} in this framework is simply some communication of the fact that the player should try to reach achievement nodes. We do not closely examine these communication mechanisms (e.g. Minecraft's achievement tree conveys a great deal of information about achievements' inter-dependency, whereas in some games, how to complete achievements is left completely opaque).

\subsection{Ceptre}
Designing quests from the rules of a game requires formalizing its mechanics in a way that is analyzable by an algorithm. We created our formal model of Minecraft's mechanics in Ceptre~\cite{ceptre2015}, which allows us to represent game logic in a high-level, quickly-prototypable, yet rigorous and formal way. 

Ceptre is a rule-based specification language describing how program states (e.g. game states) may evolve.
Each Ceptre rule consists of a left-hand side (LHS), its inputs, and a right-hand side (RHS), its outputs.
%These two sides are syntactically separated by the \verb|-o| ``lolli'' operator.
%A rule's inputs and outputs are specified as logical predicates, separated by the conjunctive ``tensor'' symbol written \verb|*|. Predicates may have arguments, which may be concrete terms declared in the program, or variables ranging over all terms of a certain type.
%
A collection of rules is interpreted by {\em multiset rewriting}: an initial state is supplied, a multiset of predicates. Then, each step of execution involves selecting a rule that may fire in the current state, and firing it, which means replacing the elements in the multiset that match the LHS of the rule with new elements matching the RHS.

Collections of rules within Ceptre programs are called {\em stages}, and each stage may be designated as autonomous or interactive.
When a stage runs manually, Ceptre displays a list of possible state transitions to the user from which she may choose to step the program forward. If the program is run autonomously, it will randomly pick possible transitions until no more remain.

Regardless of which method is used, it is possible to view a visual representation of the rules that were executed by generating the trace graph such as the one shown in Figure \ref{fig:trace}. These traces provide information about how rules fit together to help establish where any progression occurs. Being able to analyze how rules interoperate this way is the main reason Ceptre was chosen for this project.
Running Ceptre code autonomously multiple times creates a random sampling of play trace space and generates corresponding trace graphs that visually demonstrate how the rules interact, while manually running the code allows the user to see each step of the process.

%% file: model.tex
\begin{table}
\begin{tabular}{c|c}
{\bf Predicate} & {\bf Meaning}\\
\hline
\verb|tree| $N$ & There are $N$ trees in the environment\\
\verb|wood_block| $N$ & Player has $N$ wood blocks\\
\verb|cT| & Player has crafting table\\
\verb|plank| $N$ & Player has $N$ wood planks\\
\verb|stick| $N$ & Player has $N$ sticks\\
\verb|cobble| $N$ & Player has $N$ cobblestone pieces\\
... & ...\\
\verb|pickaxe| $M$ & Player has a pickaxe of material $M$\\
\verb|sword| $M$ & Player has a sword of material $M$\\
... & ...
\end{tabular}
\caption{Predicates used in our model. Predicates used in the paper are shown; additional predicates included in the formal model are indicate with ellipses.}
\label{tab:predicates}
\end{table}

The game mechanics we chose to look at in Minecraft fall into two groups, rules for gathering materials and rules to craft tools from those materials. In this section, we discuss the Minecraft rules we chose to represent and how they are formalized in Ceptre.

First, we encoded various facts about the world, such as the quantity of materials carried by the player, which tools she has built out of which materials, and so on, as predicates in Ceptre. The mapping from predicate to Minecraft state is shown in Table~\ref{tab:predicates}. Note that predicates with arguments are written in Ceptre with adjacency, e.g. \verb|pred Arg|, rather than with parentheses.

\subsection{Rules}

\begin{figure}
\centering
\includegraphics[width=0.3\textwidth]{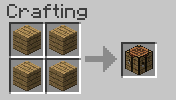}
\begin{verbatim}
craftTable : plank (N + 4) -o cT * plank N. 
\end{verbatim}
\caption{Crafting a crafting table using the inventory menu and its formalization in Ceptre.}
\label{fig:crafting}
\end{figure}

\subsubsection{Rules for Gathering}
The first category of rules is based on taking resources found in the world of Minecraft and refining them into materials for crafting items and tools. For the purpose of this project, a distinction has been made between gathering and mining. While both involve collecting resources, mining refers to the materials that can only be collected when using a pickaxe, one of the tools in the game.
Due to this limitation, the gathering rules only include the recipes needed to make a wood pickaxe and begin mining.

An example of a gathering rule is one that will allow the player to gather wood from a tree:

\begin{verbatim}
 tree : wood (N+1) -o wood_block 1 * wood N.
\end{verbatim}

How to parse this rule: the rule is named \verb|tree|. The name is separated from the rule itself with a colon (\verb|:|). The rule's LHS is \verb|wood (N+1)|, meaning that for the rule to fire, there must be some number of trees representable as $N+1$ for some $N$. After the rule fires, this predicate will be {\em consumed}, meaning there will no longer be $N+1$ trees. The rule's RHS is \verb|wood_block 1 * wood N|, which is two predicates conjoined with tensor (\verb|*|). The first predicate, \verb|wood_block 1|, means that firing the rule results in the player having 1 wooden block. The second predicate \verb|wood N| means that after the rule fires, there will be $N$ trees, which is 1 fewer than what we started with.

At an early stage in the game, the player has access to a crafting interface that is a 2x2 grid serving as the input for materials, and a single square on the right for the recipe's output. By filling the grid with wooden planks, the player is able to produce a crafting table block. Interacting with the crafting table is what provides access to the 3x3 crafting menu needed for more complex recipes, such as tools. The rule for creating a crafting table is shown in both Minecraft's crafting interface and as a rule in Ceptre in Figure~\ref{fig:crafting}.

% This rule is named \verb|craftTable|. Its uses the predicates \verb|plank| for wooden planks and \verb|cT| for the crafting table. The rule name and the rule itself are separated by a colon \verb|:|. 

% The fact that the rule's right-hand side is \verb|plank (N + 4)| means two things: first of all, the rule only applies in situations where the player has at least 4 planks; otherwise, the game state does not {\em pattern match} to this precondition. Second, it means that {\em when} this rule is selected to fire, the fact that the player has $N+4$ planks is {\em removed} from the game state and replaced by the right-hand side.

% The right-hand side of the rule is \verb|cT * plank N|. This is a conjunction of the facts \verb|cT| and \verb|plank N|, meaning {\em the player has a crafting table} and {\em the player has $N$ planks, respectively}. The $N$ in the second predicate refers to the same $N$ in the LHS. In other words, if the player started with some quantity of planks representable as $N+4$ for some (any) $N$, then after the rule fires, they have $N$, 4 fewer than they started with. The \verb|plank| predicate is typed such that its argument is a natural number, meaning the model can never transition to a state where the player has fewer than 0 planks.

\subsubsection{Rules for Tools and Mining}
After creating a crafting table and gaining access to the 3x3 interface, the program can move on to tools and the materials needed to make them. The tools in Minecraft are shovels, swords, axes, hoes, and pickaxes. All of these recipes involve a combination of materials and sticks. The material used to create a tool dictates the strength and durability of a tool. Some materials may only be collected by tools of a certain strength, so upgrading to a higher tier opens up new crafting opportunities for the player. While Minecraft has a system in place where the condition of a tool decreases after prolonged use until it breaks and disappears, for the purposes of this project that feature is not represented in the rules. If the program creates one of a tool at a certain strength level, it does not need to make another of that same material. The strength levels go from wood as the weakest, then stone, iron/gold, and diamond is the strongest. Gold is a special case in that it requires an iron pickaxe to be collected, but is not stronger than iron. All of these materials must be mined using a pickaxe of the tier directly below it, except for wood which is already covered under gathering.

\begin{verbbox}[\small]
craftWoodPickaxe : $cT * stick (N + 2) * plank (F + 3)
    -o pickaxe wood * (stick N) * (plank F).
\end{verbbox}
\begin{figure}[h!]
	\includegraphics[width=0.4\textwidth]{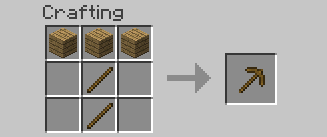}
    \theverbbox
    \caption{The recipe for crafting a wooden pickaxe in Minecraft and the Ceptre formalization of this action.}
    \label{fig:woodpick}
\end{figure}

Pickaxes in Minecraft are crafted by placing three of a material in the top row and two sticks in the remaining spots of the middle column, as shown in Figure \ref{fig:woodpick}. The Ceptre representation of this recipe is shown underneath the game screenshot. The rule consumes a crafting table, at least two sticks, and at least three wooden planks, and it produces a wooden pickaxe and any remaining sticks and planks, as well as the crafting table. The crafting table is not destroyed during this process, so the symbol ``{\$}'' is used to indicate that the predicate appears on both sides of the rule. The term \verb|wood| is the material argument to the \verb|pickaxe| predicate.

The player can then use a wooden pickaxe to mine stone, codified by the following rule:

\begin{verbatim}
mineStone : $pickaxe M * stone (N + 1) * cobble C -o 
    cobble (C + 1) * stone N.
\end{verbatim}

When given a pickaxe and at least one block of stone, this mine/stone rule can be called to return the same pickaxe, one more piece of cobblestone, and any remaining stone. The difference between cobblestone (represented by the ``cobble'' predicate) and stone in Minecraft is that stone is found more commonly, but becomes cobblestone when broken with a pickaxe. Cobblestone is used for crafting stone tools, while stone cannot. Also, breaking stone without a pickaxe results in no cobble being dropped, which is why wooden pickaxes must be crafted beforehand. However, a pickaxe of any material can be used to mine stone, so the ``M'' signifies that any pickaxe may be used as a predicate.

\begin{verbbox}[\small]
craftStonePickaxe : $cT * stick (N + 2) * cobble (F + 3)
    -o pickaxe s * (stick N) * (cobble F).
\end{verbbox}
\begin{figure}[h!]
	\includegraphics[width=0.4\textwidth]{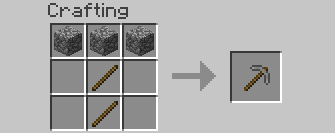}
    \theverbbox
    \caption{Crafting a stone pickaxe in Minecraft and in Ceptre.}
    \label{fig:stonepick}
\end{figure}

Once the player has stone, they are able to use it to make tools like in Figure \ref{fig:stonepick} where a stone pickaxe is being Crafted in Minecraft and in Ceptre. The similarities between Figure \ref{fig:woodpick} and Figure \ref{fig:stonepick} are evident both in the game and in the code. Any wooden planks in the recipe were replaced by cobblestone. This basic pattern became a key component in determining quest placement.

%% file: tanalysis.tex
\begin{figure}
	\includegraphics[width =\linewidth]{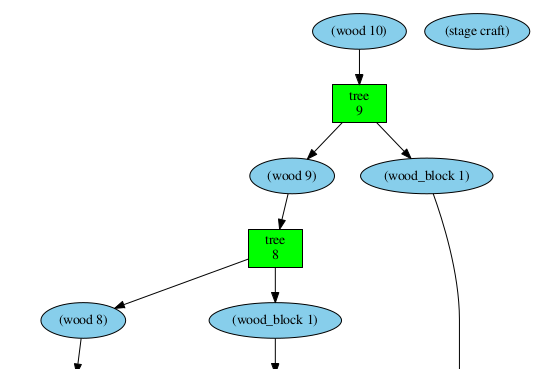}
	\caption{A portion of a trace graph. Predicate nodes are shown as blue ellipses, while transition nodes are green rectangles. An edge from a predicate to a transition indicates that the transition consumes the predicate, while an edge from a transition to a predicate suggests that the transition produces the predicate.}
	\label{fig:trace}
\end{figure}

In order to understand how the rules operate in relationship to each other, the program was run both autonomously and interactively. These procedures provided insight into how the rules were connected, allowing us to analyze the progression elements in the gameplay. When generating these models, the main patterns we looked for were bottlenecks and feedback loops. The gameplay of Minecraft, specifically the mining aspect, tends to be cyclical with players repetitively making stronger tools and finding stronger materials. Running the Ceptre model helped us identify key rules that stratify this process into stages.

As more rules fire, the number of possible choices in the system increases. Each run begins with the same rules being necessary, specifically the rules for gathering wood and making the crafting table and sticks---the crafting table recipe is an important node in all play traces, because no tools may be crafted until the player has made it. Such a node represents a {\em bottleneck} in the mechanics.

Once the program has these materials, rules for wooden tools can be run, which opens access to the mining rules. The mining rules must fire multiple times in order to have sufficient resources to create a tool. The way that these rules are repeated hints towards their significance in the player experience. This interactive model represents the player experience through text, but is similarly time-consuming. After the initial discovery of how important the mining rules were and how the crafting table expands the player's options, we moved on to studying the trace graphs to generate larger batches of data to compare.

Trace graphs are a visual representation of all rules selected during a run of the program---see Figure~\ref{fig:trace}. This trace graph shows that the \verb|tree| rule is dependent on itself, since having some number of trees in the environment is both a requirement for the rule and an outcome. We can also see that the numeric {\em argument} to the rule decreases on each subsequent firing of the rule, and that the other resources generated (the wood blocks) are used in other, unrelated rules.
% The process of the model autonomously selecting rules was generally more inefficient as it would generate multiple versions of the same tools. This inefficiency was offset by how quickly the graphs could be generated, 
%
These graphs allowed us to compare multiple possible outcomes when the program is given the same initial resources. These comparisons reinforced the results found through the interactive testing and highlighted the significance of the mining rules. The high likelihood that a player will perform these actions multiple times during a playthrough indicates that objectives such as ``Mine X amount of stone'' might be worthwhile. While these quests tend to be viewed as unoriginal and tedious, if the value for ``X'' is chosen based on the expected amount of times a player will perform this action through the course of playing the game naturally, it can potentially make a task that is tedious in general more rewarding. Creating multiples of these trace graphs and analyzing how many times certain rules are run can help determine what an appropriate ``X'' value is for that rule.

%% file: sanalysis.tex
We have identified two different methods for organizing rules that separate segments of a trace into different phases base on static analysis of the LHS for each rule. We generated a static dependency graph between rules, where an edge between $R_1$ and $R_2$ indicates that $R_2$ consumes predicates that $R_1$ produces. Then, we collapsed rules with similar LHS together into a single node with a list of possible alternate arguments. We then interpret the bottleneck points and actions immediately following them as game events that should be rewarded with achievements.
The two factors that were used to determine which LHSes were similar were the {\em materials} represented as arguments to predicates and the {\em quantities} of raw materials that a rule required, which corresponds to its tool type.

\begin{figure}[h]
	\includegraphics[width = \linewidth]{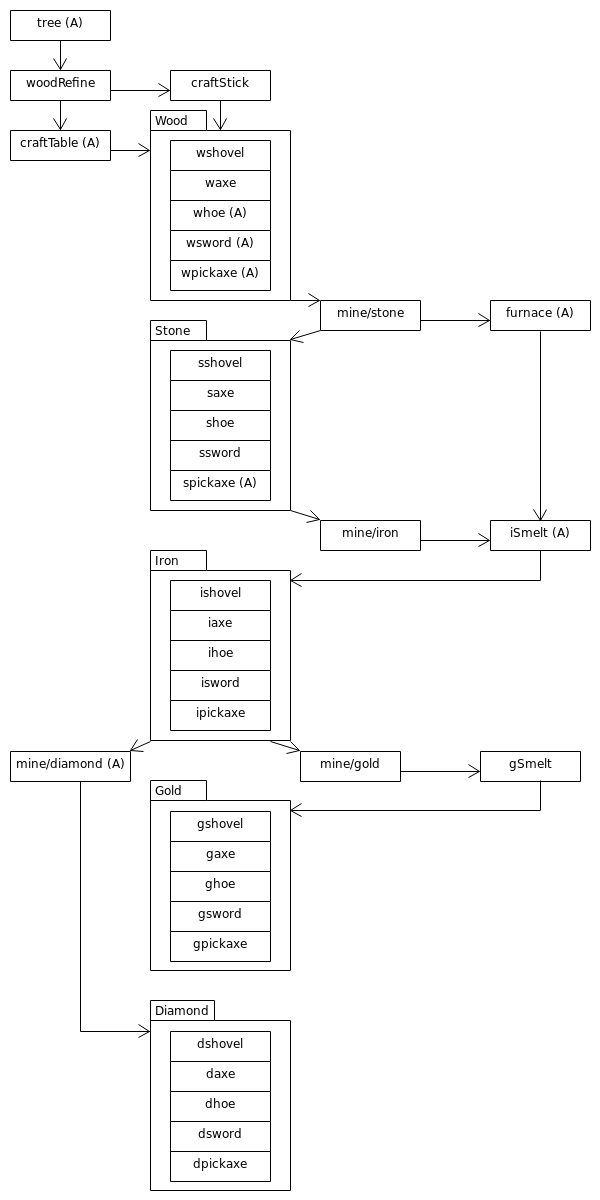}
    \caption{A visualization of the tool rules organized based on the type of materials used. (A) marks where Minecraft currently has achievements.}
    \label{fig:matGroup}
\end{figure}
\begin{figure*}
	\includegraphics[width=\textwidth]{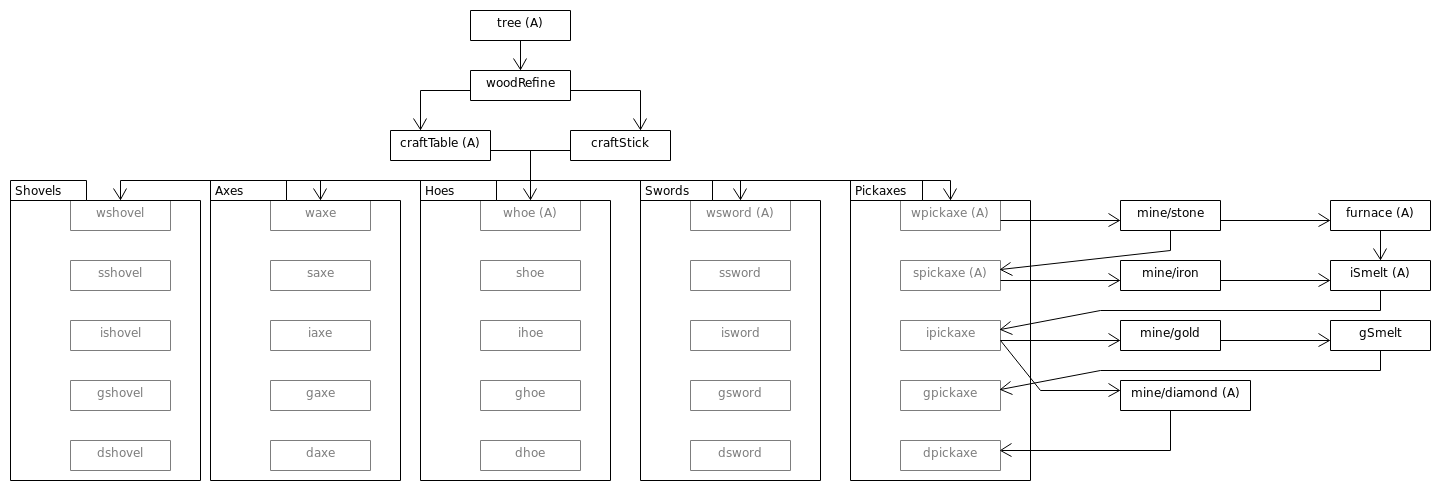}
    \caption{The tool rules categorized based on the type of tool that is generated. (A) marks where Minecraft currently has achievements.}
    \label{fig:typeGroup}
\end{figure*}
\paragraph{Material-Based Sorting}
One way to sort these rules is based on the type of material they require. All tools in Minecraft are a combination of sticks and the material that determines the tools strength. Figure \ref{fig:matGroup} groups rules that have the same inputs together. The wood section is all of the rules that take some amount of sticks and wood as inputs, the stone rules have inputs of sticks and stone and so on. Objectives based on this type of organization could reward players for unlocking improved versions of items or abilities that were previously accessible. An example of this is the Minecraft achievement "Getting an Upgrade" that rewards the player for acquiring a stone pickaxe, which in Figure \ref{fig:matGroup} would be the seen as moving from the wood group to the stone group. These types of goals reward the player by enhancing an aspect of the gameplay that they are already familiar with. In Minecraft, the stone pickaxe performs the same basic mining function as the wooden one, but does it faster and on resources that are too strong to be collected by the wood pickaxe. This method of organizing is useful for finding objectives that make the player stronger.

\paragraph{Quantity-Based Sorting}
A different approach to organizing rules is based on the number of materials that the rule takes. In Minecraft, pickaxes are made with two sticks and three of another material, while swords are one stick and two of a material. Focusing on the format of the recipe instead of on the specific ingredients leads to grouping tools that have similar functionality , but differing strength. This method is shown in Figure~\ref{fig:typeGroup} where rules are grouped by the type of tool that is produced. There are achievements for the player crafting their first hoe, sword, and pickaxe, which are all marked in the first node of their respective groups. These achievements all represent the player unlocking a new element of gameplay. Hoes, swords, and pickaxes are the critical tools for farming, combat, and mining respectively. Categorizing rules based on the quantities of inputs aids in the formation of objectives that reward the player with new ways to play the game.

%% file: conclusion.tex
Through our formalization and analysis of the rules of Minecraft in Ceptre, we have shown that it is possible to re-discover hand-authored progression structures algorithmically from the open-world mechanics of the game.
This process can help explain and identify ideal achievement placement in open world settings. We carried out two analyses: an informal examination of the play traces resulting from autonomous execution of the rules, and a formal analysis of the static, syntactic structure of the encoded rules themselves. The algorithmic discoveries we have made about the system dynamics match up with achievements that were manually authored for Minecraft. 

In future work, we hope to apply this process of deriving progression structures from mechanics to other games, particularly those that do not already have manually-authored quests or achievements. We also intend to devise alternative achievement placement strategies and compare them to the two sorting-based strategies.

The ultimate goal of such a research programme is to improve the player's sense of purpose in open world games by providing more affordances for potential intermittent objectives, where those objectives stem organically from the gameplay itself.